\documentclass[aps, prb, twocolumn, showpacs,superscriptaddress]{revtex4-1}
\usepackage{setspace}
\usepackage{dcolumn}
\usepackage{graphicx}
\usepackage{bm, amsmath, amssymb, textcomp}
\usepackage[colorlinks=true,linkcolor=blue,citecolor=blue,urlcolor=blue]{hyperref}
\usepackage{color,soul}

\newcommand{\xz}[1]{\textcolor{red}{{Xiao: #1 }}}

\begin{document}

\title{Ab initio theory of free-carrier absorption in semiconductors}

\author{Xiao Zhang}
\affiliation{Department of Materials Science and Engineering, University of Michigan, Ann Arbor, Michigan, 48109, USA}
\author{Guangsha Shi}
\affiliation{Department of Materials Science and Engineering, University of Michigan, Ann Arbor, Michigan, 48109, USA}
\author{Joshua A. Leveillee}
\affiliation{Oden Institute for Computational Engineering and Sciences, University of Texas, Austin, Texas, 78712, USA}
\author{Feliciano Giustino}
\affiliation{Oden Institute for Computational Engineering and Sciences, University of Texas, Austin, Texas, 78712, USA}
\author{Emmanouil Kioupakis}
\affiliation{Department of Materials Science and Engineering, University of Michigan, Ann Arbor, Michigan, 48109, USA}
\date{\today}

\begin{abstract}
The absorption of light by free carriers in semiconductors results in optical loss for all photon wavelengths.
Since free-carrier absorption competes with optical transitions across the band gap, it also reduces the efficiency of optoelectronic devices such as solar cells because it does not
generate electron-hole pairs.
In this work, we develop a first-principles theory of free-carrier absorption taking into account both single-particle excitations and the collective Drude term, and we demonstrate its application to the case of doped Si.
We determine the free-carrier absorption coefficient as a function of carrier concentration and we obtain excellent agreement with experimental data.
We identify the dominant processes that contribute to free-carrier absorption at various photon wavelengths, and analyze the results to evaluate the impact of this loss mechanism on the efficiency of Si-based optoelectronic devices.


\end{abstract}

\maketitle



\section{Introduction}
Optical absorption across the band gap of indirect-gap semiconductors can only occur if accompanied by momentum transfer via, e.g., electron-phonon scattering. 
Phonon-assisted indirect absorption is especially important for silicon (Si), the most widely deployed semiconductor for electronic and solar-cell devices.
Due to the need of considering second-order phonon-assisted processes, the computational cost of studying indirect absorption is much higher compared to direct absorption, and theoretical approaches that can address the problem have only emerged in recent years\cite{Noffsinger2012,zacharias2016one,zacharias2020theory}. 
In particular, the combination of density functional perturbation theory (DFPT)\cite{RevModPhys.73.515} with maximally localized Wannier functions (MLWF)\cite{RevModPhys.84.1419} allow the exploration of electron-phonon interactions and indirect optical absorption with ultra-fine sampling of the Brillouin zone (BZ)\cite{Noffsinger2012}. 

Apart from the cross-gap absorption, it is also important to consider optical absorption introduced by free carriers in doped semiconductors, particularly for photon energies below the band gap. 
For example, Si photovoltaic devices always involve n-type and p-type doped regions to create a p-n junction and enable charge separation. 
The free carriers in these regions act as additional sources of optical absorption, especially in the infrared regime.
Traditionally, free-carrier absorption (FCA) is accounted for with the semi-classical Drude model, which describes the energy absorption of the light field induced by electrical resistivity due to the collective oscillation of the carriers.
However, at relatively short wavelengths including the visible
, near-IR
, short- 
and mid-wavelength IR regions
, the semi-classical Drude model becomes unreliable to fully describe FCA. 
In these regions, quantum-mechanical first-principles methods are necessary as demonstrated previously for nitrides and transparent-conducting oxides\cite{PhysRevB.81.241201,PhysRevB.92.235201,PhysRevB.100.081202}. 

In this work, we demonstrate a general first-principles theory to study FCA in order to understand the optical properties in the infrared region for doped semiconductors.
We explore the following four different FCA mechanisms: 1) direct absorption between the lowest conduction bands or highest valence bands; 2) phonon-assisted absorption\cite{Noffsinger2012}; 3) charged-impurity-assisted absorption, and 4) the semi-classical resistive Drude term. 
We demonstrate the application of the method on both n-type and p-type Si.
By comparing the calculated absorption coefficient to experiments\cite{PhysRev.108.268, hara1966free}, we find excellent agreement across a wide range of carrier densities for photon wavelengths from 1 $\mu m$ to 10 $\mu m$. 
We also demonstrate that in the near to mid-IR region, the direct, phonon-assisted, and resistive terms all play an important role and none can be neglected. 
Importantly, we also demonstrate that FCA in heavily doped Si is comparable to cross-gap absorption for photon energies near its fundamental band gap, thus free carriers serve as a strong optical-loss mechanism. 
We demonstrate that free-carrier-induced optical loss can significantly affect applications of Si in the IR region such as thermal dectors and emitters\cite{SCLAR1984149, o2015silicon,lin2018mid}, night vision\cite{chrzanowski2013review,huang2020black}, astronomy applications\cite{marsh2007production}, thermal energy storage systems\cite{meroueh2020thermal}, etc. 

\section{Theory and Computational Approaches}

\begin{figure}
\includegraphics[width=0.95\columnwidth]{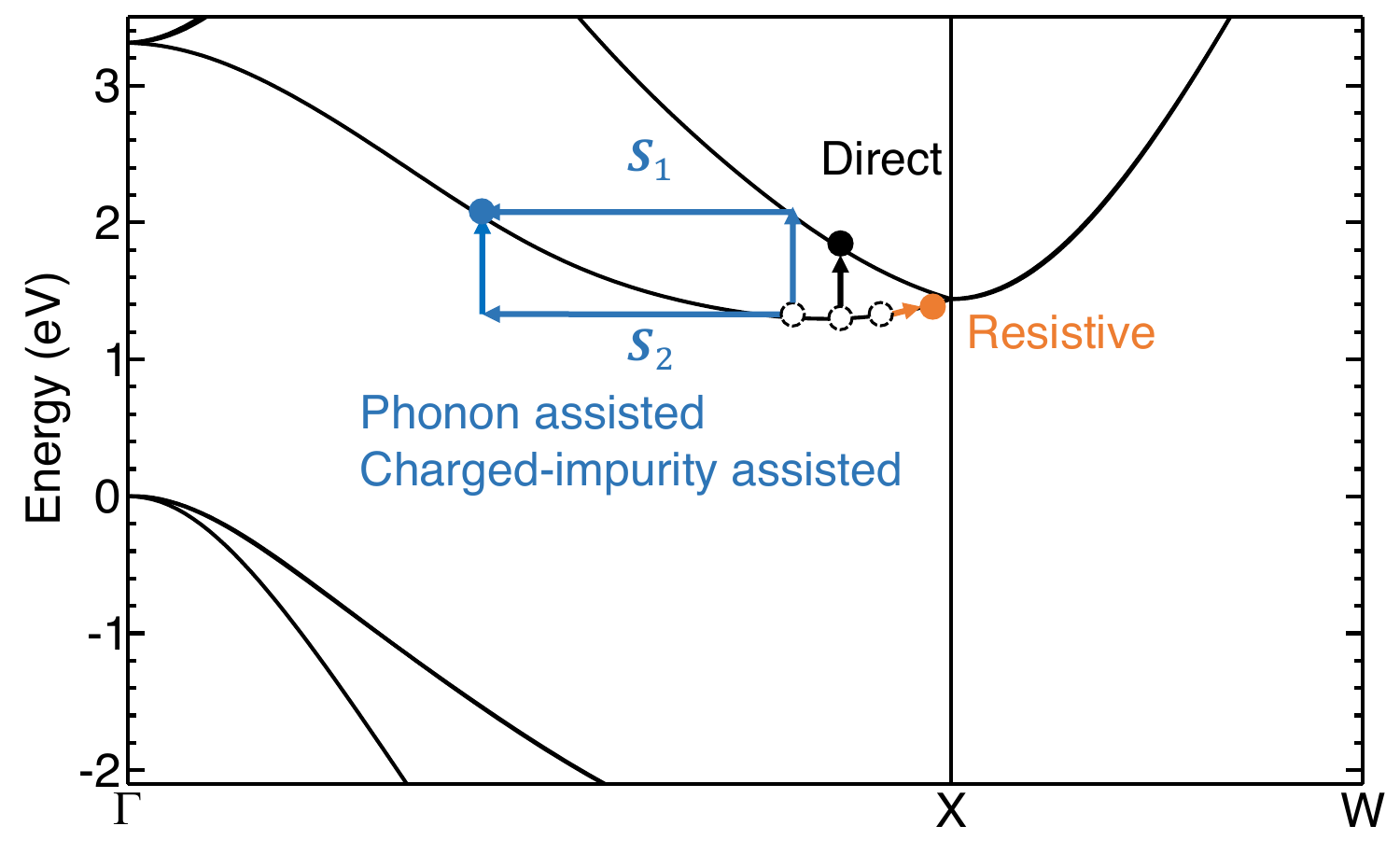}
\vspace{-0.2in}
\caption{\label{fig:bands}(Color online.) Quasiparticle band structure of Si depicting the different FCA mechanisms: direct, phonon- or charged-impurity-assisted FCA, and the resistive term for free electrons. Indirect absorption can proceed along two paths associated with generalized optical matrix elements $\mathbf{S}_1$ and $\mathbf{S}_2$.
FCA processes for free holes are similar but instead occur close to the valence band maximum at $\Gamma$. 
}
\vspace{-0.27in}
\end{figure}

The various contributions to FCA in Si are illustrated on the quasiparticle band structure in Fig.~\ref{fig:bands}.
When a free electron at the bottom of the conduction band absorbs a photon, several possible transitions may occur:
First, the electron can undergo a direct vertical transition to a higher conduction band via energy transfer from the photon.
Second, the electron can undergo an indirect transition with the additional required momentum transfer supplied by a carrier-scattering process.
The scattering mechanisms considered in this work are those mediated by phonons or by ionized impurities. 
The two possible paths of the indirect absorption are illustrated in Fig.\ref{fig:bands} as $\mathbf{S}_1$ (electron-photon followed by electron-phonon or electron-charged-impurity coupling) and $\mathbf{S}_2$ (electron-phonon or electron-charged-impurity followed by electron-photon coupling).
In addition, the finite electrical resistivity of the material provides an additional intraband loss route, which is the semi-classical Drude contribution.

The imaginary part of the dielectric function resulting from direct and indirect (phonon- or charged-impurity-assisted) absorption are calculated via standard Fermi's golden rule and second-order time-dependent perturbation theory, respectively.
For direct absorption, the imaginary part of the dielectric function is obtained by\cite{PhysRevB.92.235201}:
\begin{align}\label{eq:direct_epsilon2}
\text{Im }\varepsilon^{\text{dir}}(\omega)=\frac{8\pi^2e^2}{V \omega^2}\frac{1}{N_{\mathbf{k}}}\sum_{ij\mathbf{k}} & \left(f_{j\mathbf{k}}-f_{i\mathbf{k}}\right) \left|\mathbf{e}\cdot\mathbf{v}_{ij}(\mathbf{k})\right|^2 \nonumber \\
&\times \delta(\epsilon_{j\mathbf{k}}-\epsilon_{i\mathbf{k}}-\hbar\omega).
\end{align}
In Eq.(\ref{eq:direct_epsilon2}), an extra pre-factor of two is from considering electronic spin degeneracy. $V$ is the volume of the unit cell. $N_{\mathbf{k}}$ is the number of $\mathbf{k}$-points for the Brillouin zone (BZ) sampling. 
$\mathbf{v}_{ij}(\mathbf{k})=\langle j\mathbf{k}|\mathbf{v}|i\mathbf{k}\rangle$ is the velocity matrix element. 
$\mathbf{e}$ is the direction of the polarization of light, and $\hbar\omega$ is the photon energy. Indices $i$ and $j$ indicate the initial and final band indices. $\mathbf{k}$ is the electron wave vector, and $\epsilon_{i\mathbf{k}}$ and $f_{i\mathbf{k}}$ are the energy and Fermi-Dirac occupation number of state $(i,{\mathbf{k}})$. For the phonon-assisted process, the imaginary part of the dielectric function is expressed as\cite{Noffsinger2012,PhysRevB.92.235201}:
\begin{align}\label{eq:phonon_epsilon2}
\text{Im }\varepsilon^{\text{ph}}(\omega)=& \frac{8\pi^2 e^2}{V \omega^2}\frac{1}{N_{\mathbf{k}}N_{\mathbf{q}}} \nonumber\\ &\times\sum_{\nu ij\mathbf{kq}} \left|\mathbf{e}\cdot(\mathbf{S}_{1,ij\nu}(\mathbf{k,q})+\mathbf{S}_{2,ij\nu}(\mathbf{k,q}))\right|^2 \nonumber \\ 
& \times P_{ij\nu}(\mathbf{k,q})\delta(\epsilon_{j\mathbf{k}+\mathbf{q}}-\epsilon_{i\mathbf{k}}-\hbar\omega\pm\hbar\omega_{\nu\mathbf{q}}).
\end{align}
In Eq.(\ref{eq:phonon_epsilon2}), $\mathbf{q}$ is the wave-vector of the phonons and $N_{\mathbf{q}}$ is the number of $\mathbf{q}$-points for the BZ-sampling of the phonons. $\omega_{\nu\mathbf{q}}$ is the frequency of the phonon mode $\nu$ at $\mathbf{q}$. The phonon-assisted absorption contains two possible paths illustrated by $\mathbf{S}_1$ and $\mathbf{S}_2$ in Fig.\ref{fig:bands}. 
The generalized matrix elements for the two paths of phonon assisted absorption are given by:
\begin{align}\label{eq:indirect_S1}
\mathbf{S}_{1,ij\nu}(\mathbf{k}, \mathbf{q})=\sum_m\frac{\mathbf{v}_{im}(\mathbf{k})g_{mj,\nu}^\text{el-ph}(\mathbf{k},\mathbf{q})}{\epsilon_{m\mathbf{k}}-\epsilon_{i\mathbf{k}}-\hbar\omega+i\eta}
\end{align}
and
\begin{align}\label{eq:indirect_S2}
\mathbf{S}_{2,ij\nu}(\mathbf{k}, \mathbf{q})=\sum_m\frac{g_{im,\nu}^\text{el-ph}(\mathbf{k},\mathbf{q})\mathbf{v}_{mj}(\mathbf{k}+\mathbf{q})}{\epsilon_{m\mathbf{k}+\mathbf{q}}-\epsilon_{i\mathbf{k}}\pm\hbar\omega_{\nu\mathbf{q}}+i\eta},
\end{align}
 where $g_{ij,\nu}^\text{el-ph}(\mathbf{k},\mathbf{q})$ are the electron-phonon coupling matrix elements between electron states $(i,\mathbf{k})$ and $(j,\mathbf{k}+\mathbf{q})$ through phonon mode $(\nu \mathbf{q})$. 
$\eta$ is a numerical broadening parameter to avoid the divergence in the denominator. 
The factor $P_{ij}$ is given by considering detailed balance for the phonon-absorption [$P_{a,ij\nu}(\mathbf{k,q})$] process: 
\begin{equation}
\label{eqn:pa}
\begin{aligned}
    P_{a,ij\nu}(\mathbf{k,q})=&n_{\mathbf{\nu\mathbf{q}}}\times f_{i\mathbf{k}}\times(1-f_{j\mathbf{k+q}})\\&-(n_{\nu\mathbf{q}}+1)\times(1-f_{i\mathbf{k}})\times f_{j\mathbf{k+q}}
\end{aligned}
\end{equation}
and the phonon-emission [$P_{e,ij\nu}(\mathbf{k,q})$] process:
\begin{equation}
\label{eqn:pe}
\begin{aligned}
    P_{e,ij\nu}(\mathbf{k,q})=&(n_{\mathbf{\nu\mathbf{q}}}+1)\times f_{i\mathbf{k}}\times(1-f_{j\mathbf{k+q}})\\&-n_{\nu\mathbf{q}}\times(1-f_{i\mathbf{k}})\times f_{j\mathbf{k+q}}.
\end{aligned}
\end{equation}
where $n_{\nu\mathbf{q}}$ is the phonon Bose-Einstein occupation number. 
The expressions for the charged-impurity-assisted terms are similar to that for the phonon-assisted absorption with the following difference:
For charged-impurity scattering, the occupation factor is the same as in direct absorption. In addition, the sum over phonon modes and the phonon frequencies are not present, and the matrix elements replaced by the electron-charged-impurity matrix elements discussed below.  
The carrier-charged-impurity scattering process is modeled by considering the screened Coulomb interaction between the free carrier and the charged impurities. The matrix elements are given by:
\begin{equation}
\begin{aligned}\label{eq:g_impurity}
g_{ij}^\text{imp}(\mathbf{k},\mathbf{q})=&\frac{4\pi e^2Z}{V}\\&\times\sum_{\mathbf{G}\neq\mathbf{-q}}\langle j\mathbf{k+q}|\frac{e^{i(\mathbf{q+G})\cdot\mathbf{r}}}{\varepsilon_0(|\mathbf{q+G}|^2+q_D^2)}|i\mathbf{k}\rangle,
\end{aligned}
\end{equation}
$\mathbf{G}$ is the first shell of reciprocal lattice vectors around $\Gamma$, and $q_D$ is the screening wave vector given by the Debye model ($q_D^2=\frac{4\pi e^2 n}{\varepsilon_0 k_B T}$). 
In this work, we verified that the free-carrier screening is well-described by the non-degenerate approximation of the Debye model due to that the Fermi level is below the conduction band minimum for free electrons and above the valence band maximum for free holes.
The dopants are assumed to be singly charged ($Z=1$), which holds true for Si doped with III/V elements\cite{jacoboni1977review,PhysRevB.23.5531,PhysRev.108.268,schroder1978free}. 
In our work, the impurity is assumed to be a point charge, therefore the above formalism is most suitable to describe small-$\mathbf{q}$ (i.e., long-range) scattering. 
FCA is the main absorption mechanism in the IR region below the indirect band gap, and is dominated by small $\mathbf{q}$-vectors due to the low photon energies. 
As a result, we do not consider the $q$-dependent screening in our calculations. 
Instead, we use the dielectric constant of $\varepsilon_0=13.3$ calculated from DFPT. 
The calculated value agrees well with Ref.\cite{PhysRevB.97.121201}, and overestimates experimental measurements\cite{sze2021physics} by about 10\%.
However, we show that the contribution from charged-impurity scattering is rather insignificant compared to the other contributions. 
In addition, Eq.(\ref{eq:g_impurity}) requires the knowledge of the electronic wave function to calculate the matrix elements. 
To carry out the calculation on the fine electronic $\mathbf{k}$-grid, we approximate the overlap of the wave functions by the product of the rotation matrices in Wannier space in the small $\mathbf{q+G}$ limit: $\langle\psi_{j\mathbf{k+q}}|e^{i(\mathbf{q+G})\cdot\mathbf{r}}|\psi_{i\mathbf{k}}\rangle\approx[U_{\mathbf{k+q}}U_{\mathbf{k}}]_{ij}$\cite{PhysRevB.65.035109, RevModPhys.84.1419, RevModPhys.89.015003, PhysRevLett.115.176401}. 
Lastly, the contribution from the resistive Drude term is calculated using the semi-classical linearized Boltzmann Equation in the relaxation time approximation\cite{doi:10.1021/acsnano.5b06199}:
\begin{align}\label{eq:linearized_boltzmann_equation}
\text{Im }\varepsilon^{\text{resis}}(\omega)=\frac{4\pi\sigma}{\omega(1+\omega^2\tau^2)},
\end{align}
where $\sigma$ is the DC-conductivity and $\tau$ is the relaxation time of carriers obtained from first-principles calculations as described below.


We studied the electronic properties of Si using first-principles calculations (See detailed parameters in appendix Sec.\ref{sec:app_a}) based on density functional and many-body perturbation theory. 
The ground state properties are calculated using the Quantum Espresso (QE) package\cite{giannozzi2009quantum,giannozzi2017advanced} with the SG15 Optimized Norm Conserving Vanderbilt (ONCV) pseudopotentials\cite{Hamann13optimized,SCHLIPF201536}.
The quasiparticle band structure of Si was calculated using the one-shot GW method implemented in the BerkeleyGW code\cite{DESLIPPE20121269,hybertsen1986}. 
The phonon dispersion and electron-phonon coupling matrix elements were calculated with DFPT\cite{RevModPhys.73.515} implemented in the QE package.
Subsequently, we interpolated the quasiparticle energies and electron-phonon matrix elements with the maximally localized Wannier function method\cite{RevModPhys.84.1419} onto a fine sampling grid to ensure converged spectra.
The Wannier interpolation onto the fine grid as well as the subsequent calculations of the optical properties on the fine grid are implemented in the EPW code\cite{Giustino2007,Ponce2016,Noffsinger2012}. 
We assumed fully ionized singly-charged donors and acceptors, thus the charged-impurity density is the same as the free-carrier density.  
The relaxation time in the resistive term is determined by calculating the mobility in the constant-relaxation-time approximation and fitting 
the value of the constant relaxation time to match the calculated mobility as a function of carrier concentration from the full iterative solution of the Boltzmann transport equation (See appendix Sec.\ref{sec:app_a} for details of the calculation.). 
The absorption coefficient is subsequently evaluated with the calculated imaginary part of the dielectric function along with a constant refractive index approximation ($n_r=3.4$, from experimental measurements\cite{doi:10.1063/1.555624}). 

\section{Results and Discussions}

\begin{figure}
\includegraphics[width=\columnwidth]{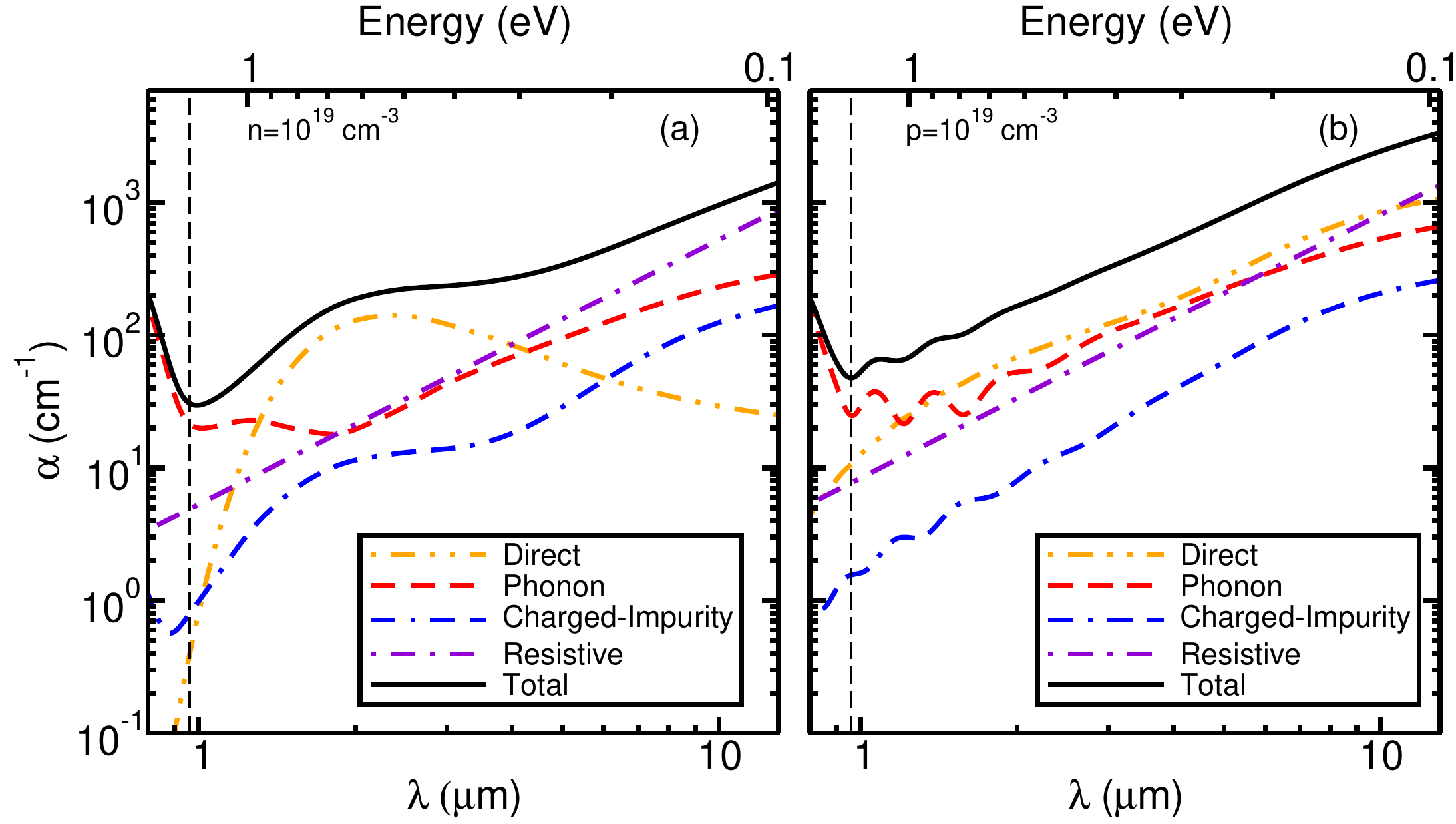}
\vspace{-0.3in}
\caption{\label{fig:si_fca_theory}(Color online.) Contributions to the FCA coefficients of Si by (a) free electrons with a concentration ($n$) of $10^{19}\text{ cm}^{-3}$, and (b) free holes with a concentration ($p$) of $10^{19}\text{ cm}^{-3}$ at 300 K. 
In both cases multiple mechanisms contribute to absorption in the near IR region. 
The vertical dashed line indicates the calculated band gap of Si ($E_g=1.29$ eV).
}
\vspace{-0.1in}
\end{figure}

We first analyze the four different contributions to FCA in n-type Si for a fixed electron concentration of $n=10^{19}\text{ cm}^{-3}$. (Fig.\ref{fig:si_fca_theory}(a))
For free electrons, the direct contribution 
shows a peak around a photon wavelength of 2 $\mu$m. 
This peak results from direct transitions from the conduction band minimum to the second lowest conduction band, whose energies differ by approximately 0.6 eV. 
The phonon-assisted contribution 
is dominated by cross-gap inter-band contribution at wavelengths shorter than the indirect band gap ($\lambda<0.96$ $\mu$m), and by both intra- and inter-conduction-band contributions at wavelengths close to and longer than the indirect band gap. 
The charged-impurity contribution 
shows a similar but much lower shoulder as the direct contribution at $\lambda=2$ $\mu$m, and increases sharply at long wavelength due to the long-range Coulomb interaction. 
The resistive contribution 
is an intra-band contribution induced by the collective oscillation of free electrons at the conduction band minimum and increases with almost a constant power as a function of wavelength ($\alpha\propto\lambda^2$).
Comparing the different mechanisms, the contributions from charged-impurity scatterings are found to be smaller by at least one order of magnitude than the dominating contributions across the entire wavelength range.
For $\lambda\le 1$ $\mu m$, the phonon-assisted contribution is the dominant term. 
This is mainly due to the calculated indirect band gap of 1.29 eV.
In this region, direct transitions are forbidden and the resistive contribution becomes small due to the $\omega^3$ term in the denominator. 
For 1 $\mu m<\lambda<4$ $\mu m$, the direct contribution dominates and induces a clear shoulder in the total absorption. 
As the wavelength further increases, at $\lambda\ge 4$ $\mu m$ the direct contribution becomes forbidden again, and only affected by the tails of the peaks around 2 $\mu m$, while the resistive contribution starts to dominate. 

Next, we analyze the different contributions in p-type Si for a fixed carrier concentration of $p=10^{19}\text{ cm}^{-3}$ (Fig.\ref{fig:si_fca_theory}(b)).
The overall features of the absorption coefficients for free holes are similar to those for free electrons. 
A notable difference is that no clear peak is observed for the direct contribution of free-hole absorption. 
This is due to the degeneracy at the valence band maximum so that the energy differences between the first and the second highest valence band continuously increase from zero away from the $\Gamma$-point. 
Comparing the different mechanisms, while the charged-impurity-assisted contribution is small across the wavelength range analyzed, the contributions from the direct, phonon-assisted, and resistive terms remain close for $1$ $\mu m<\lambda<10$ $\mu m$.
The resistive term starts to become dominant for $\lambda>10$ $\mu m$. 

\begin{figure}
\includegraphics[width=\columnwidth]{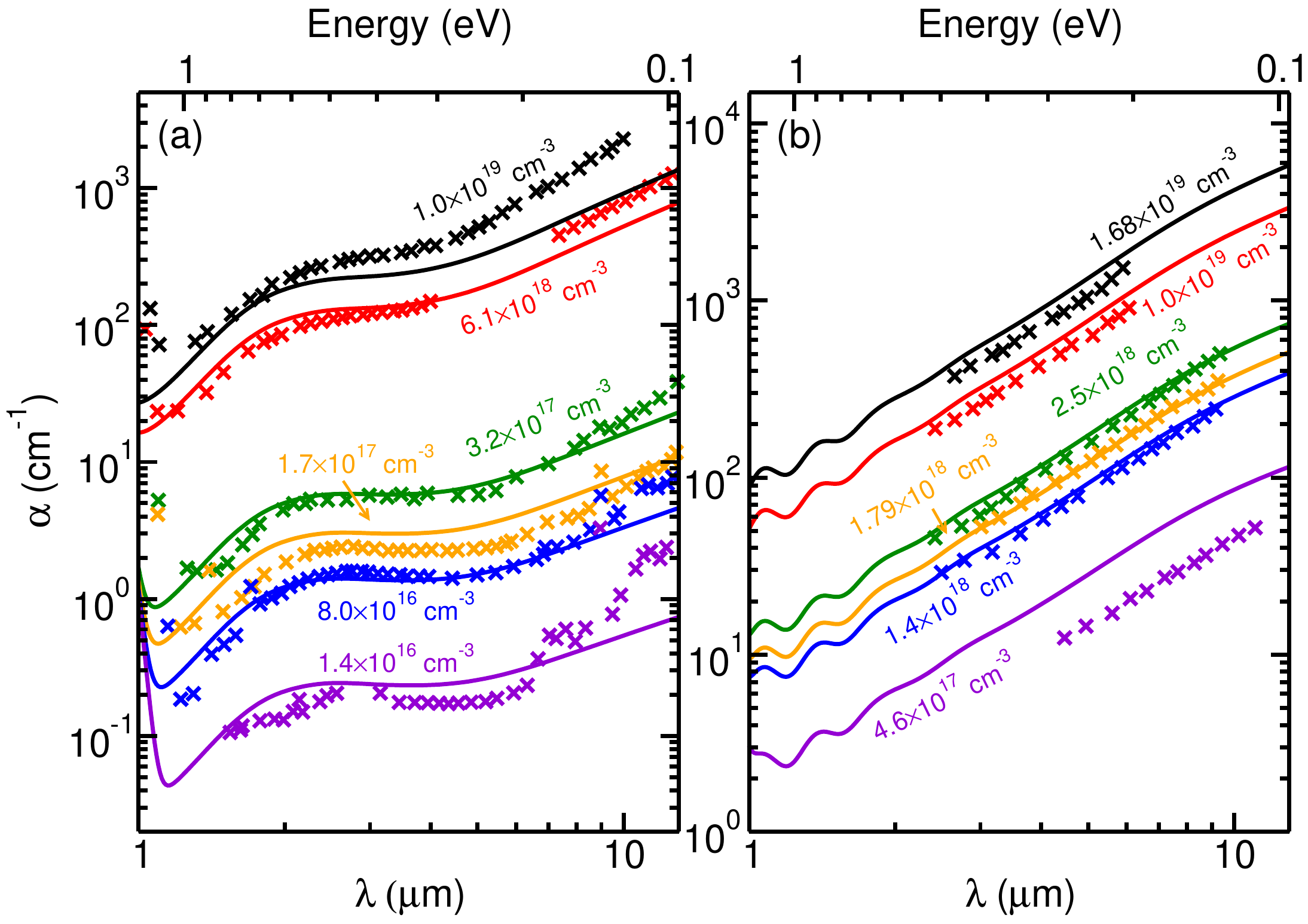}
\vspace{-0.3in}
\caption{\label{fig:compare_expt}(Color online.) Calculated FCA coefficients of (a) n-type and (b) p-type Si as a function of wavelength and carrier concentration at 300 K. 
Comparison with experiments for n-type\cite{PhysRev.108.268} and p-type\cite{hara1966free} show good agreement over a broad range of wavelengths and free-carrier concentrations. 
}
\vspace{-0.1in}
\end{figure}

The trends we find from analyzing the different contributions to FCA are important in explaining experimental observations of the optical absorption spectra of Si in the infrared region. 
We compare the combined calculated absorption coefficients to the experimentally measured values\cite{PhysRev.108.268,hara1966free} in Fig.\ref{fig:compare_expt}.
Our calculated results are in overall good agreement with the experimental measurements over a broad range of carrier concentrations ($n=1.4\times10^{16}-10^{19}$ cm$^{-3}$ and $p=4.6\times10^{17}-1.68\times10^{19}$ cm$^{-3}$) and over the wavelength range from 1 $\mu m$ to 10 $\mu m$.
Several deviation of the theoretical calculations from experimental measurements are observed and explained below: 
First, at low carrier concentrations, the sudden increase of the absorption coefficient at wavelengths longer than 6 $\mu$m is not captured by theoretical calculations. 
It has been reported before that the increase is due to lattice absorption, i.e. interaction between multiple phonons and photons\cite{PhysRev.93.674,PhysRev.108.268,johnson1959lattice,PhysRevB.69.045205}, which is not considered in our theoretical calculations focusing on free-carrier absorptions.
In addition, underestimations of the magnitude of the absorption at wavelengths longer than 7 $\mu$m for the highest two free-electron concentrations are due to the overestimation of the electron mobility by our calculations compared to experiment at these carrier concentrations (See appendix Sec.\ref{sec:app_b} for a more detailed analysis). 
Despite of the above deviations, our theoretical calculations provide crucial explanation to the dominant contributions of free-carrier absorption at different wavelengths:
For free electrons, the different ranges where different terms dominate can be clearly seen especially from the increase close to 1 $\mu$m as the wavelength decreases due to the phonon-assisted cross-gap contribution, the shoulder around 2 $\mu m$ due to the direct contribution, and the constant slope above 5 $\mu m$ due to the resistive contribution.
Importantly, for free holes, although the slopes of the curves appear to be close to constant across the range shown, the absorption results from three comparable contributions (phonon-assisted, resistive, and direct).
For both free electrons and free holes, while the resistive contribution dominates in the long-wavelength IR region ($\lambda>10$ $\mu m$), the importance of including the direct absorption and phonon-assisted absorption can be clearly seen especially in the near/short-wavelength IR region. 
\begin{figure}
\includegraphics[width=\columnwidth]{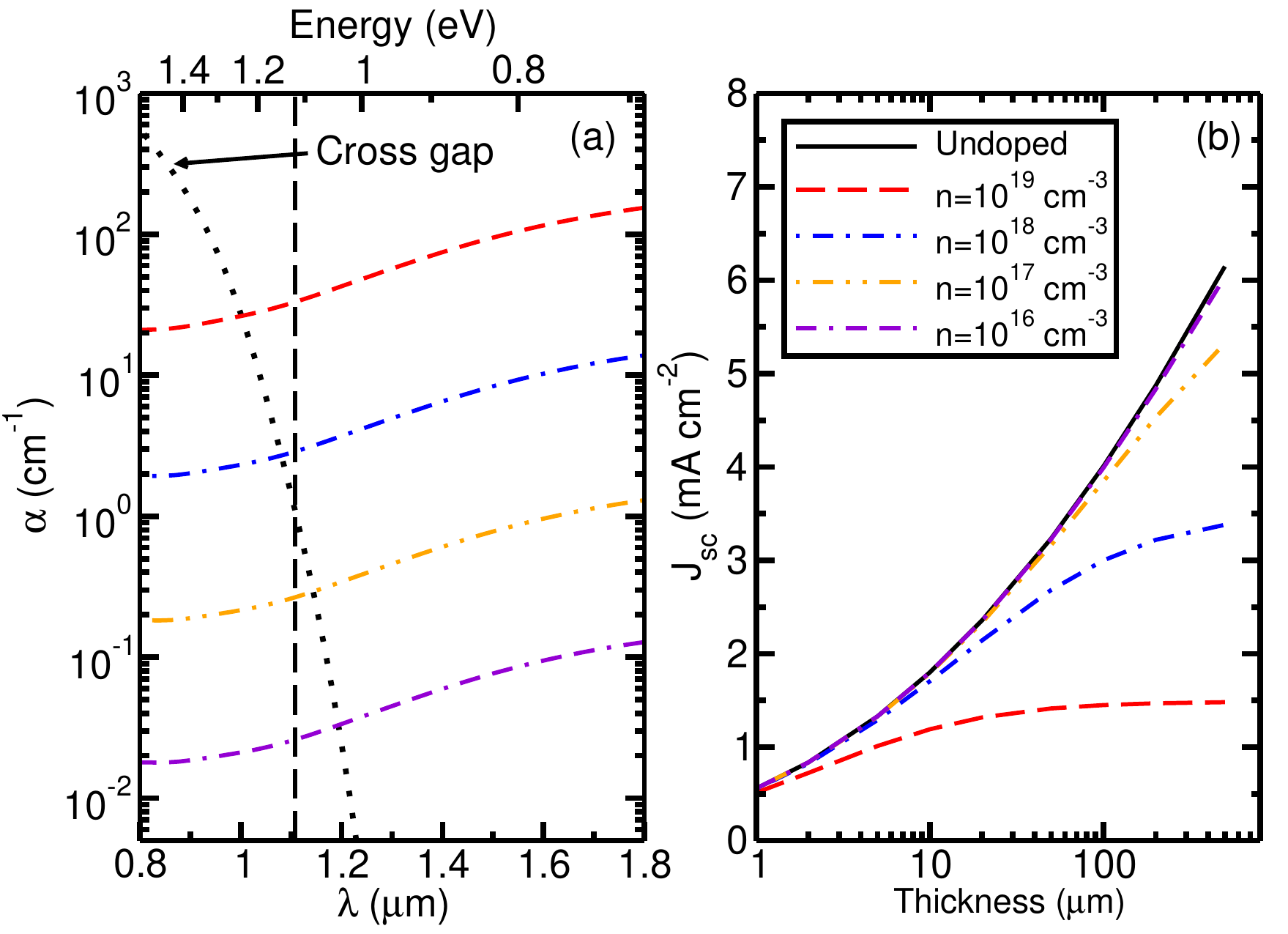}
\vspace{-0.2in}
\caption{
\label{fig:si_energy_loss}
(Color online.) (a)Calculated absorption coefficient due to FCA at different electron concentrations compared to that calculated from cross-gap absorption. 
The calculated cross-gap absorption spectrum is shifted down by 0.17 eV to match the calculated indirect band gap ($E_g$=1.29 eV) to experimental value($E^{\text{expt}}_g=1.12$ eV, dashed vertical line) at 300 K. 
(b)Short-circuit current density from electron-hole pair generations as a function of film thickness for textured films under ideal black-body radiation at 1000 K. The current density is obtained using the model illustrated by Ref.\onlinecite{tiedje1984}. 
For heavily-doped Si and thick films, free carriers induce significant optical loss in infrared optoelectronic applications of Si.  
}
\vspace{-0.1in}
\end{figure}

An important way FCA impacts practical applications is that by competing with cross-gap absorption while not generating electron-hole pairs, it can serve as a source of optical loss in optoelectronic devices.
Our comparison to experiment focused on photon energies below the band gap of Si, as it is difficult to distinguish the contribution of FCA
from the cross-gap absorption in experiment for photon energies close or larger than the band gap.
However, it is important to predict its contribution at photon wavelengths shorter than the band gap and thus better understand its contribution to energy loss in practical applications.
Fig.~\ref{fig:si_energy_loss}(a) compares the calculated FCA coefficients at various electron concentrations to that of the cross-gap absorption obtained with the same computational parameters (the cross-gap spectrum is rigidly redshifted by 0.17 eV to match the experimental band gap).
We find that FCA is much weaker than cross-gap absorption at wavelengths in the UV and visible range, but at high doping concentrations it becomes comparable to or even larger than cross-gap absorption in the near IR region (1000-1200 nm), for which photon energies become comparable to the band gap of Si.
On a practical level, however, the loss induced by free carriers is strongly dependent on the relevant spectral range of the radiation as well as the dimension of the device.
We show next that utilizing our first-principles simulation results allows us to quantify free-carrier induced losses in practical applications.  

In order to quantify the effects of free carriers on the losses in optoelectronic devices, we calculate the short-circuit current density induced by electron-hole pair generations for textured Si films of various thicknesses and electron concentrations. 
We used the model of Tiedje et al.\cite{tiedje1984} 
to calculate the absorbance $A(\hbar\omega,L)$ as a function of photon energy ($\hbar\omega$) and film thickness ($L$) from the absorption coefficient.
The absorbance of a textured film is given by:
\begin{equation}
    A(\hbar\omega,L)=\frac{\alpha_{\text{gap}}(\hbar\omega)}{\alpha_{\text{gap}}(\hbar\omega)+\alpha_{\text{fca}}(\hbar\omega)+1/(4n_r^2L)},
\end{equation}
where $\alpha_{\text{gap}}(\hbar\omega)$ and $\alpha_{\text{fca}}(\hbar\omega)$ being the absorption coefficient due to cross gap absorption and that due to FCA, respectively. 
The term of $(4n^2_rL)^{-1}$ results from considering increased mean-free path for the light ray with nonspecular textured surfaces\cite{tiedje1984}. 
The current density per photon energy $dJ_{sc}/d\hbar\omega$ is calculated via: 
\begin{equation}
    dJ_{sc}/d\hbar\omega=eA(\hbar\omega, L)n_{ph}(\hbar\omega),
\end{equation}
Where $n_{ph}(\hbar\omega)$ is the photon flux that can be obtained through the spectrum $S(\hbar\omega)$ with $n_{ph}(\hbar\omega)=S(\hbar\omega)/\hbar\omega$. The integral of $dJ_{sc}/d\hbar\omega$ over energy gives the short-circuit current $J_{sc}(L)$ as a function of film thickness. 
In our study, the absorbance is integrated with the AM 1.5 solar radiation spectrum\cite{solar}, or the 1000 K black-body radiation spectrum to obtain the short-circuit current density $J_{sc}(L)$ resulted from electron-hole pair generation of textured films under their radiation. 
We find that for solar radiation (See appendix Sec.\ref{sec:app_c} for the analysis), free carriers play a minor role in the loss mechanisms mainly due to the very thin\cite{uprety2019photogenerated} n-type emitter and low carrier density\cite{uprety2019photogenerated} in the p-type base. 
Such a conclusion is consistent with previous analysis from experimental or empirical perspectives\cite{allen2019passivating,baker2014near,richter2013reassessment}.
On the other hand, our simulations identify that FCA is a significant source of loss for thermal applications such as photodectors and night-vision devices. 
In thermal-radiation-related applications, the spectra ranges that are important shift further into the infrared region, where FCA becomes much more important.
Fig.\ref{fig:si_energy_loss}(b) shows a significant drop in the current density due to FCA for the ideal black-body radiation at 1000 K. 
For film thickness as thin as 10 $\mu m$, free carriers induce a drop of $\sim$10\% and $\sim$50\% in the short-circuit current density from electron-hole pair generation for electron densities of $n=10^{18}\text{ cm}^{-3}$ and $n=10^{19}\text{ cm}^{-3}$, respectively.
As the film thickness increases, the loss induced by free carriers becomes progressively more severe.
This observation can be understood from the fact that for thicker films, the absorption of photons in the infrared region, dominated by free carriers, becomes stronger due to the increased mean path length of light rays in the films. 
As a result, the integrated current density is more strongly affected by free carriers.

\section{Conclusions}
In summary, we developed a method to investigate free-carrier absorption (FCA) in doped Si with predictive atomistic calculations. 
The calculated results are in good agreement with experiment over several orders of magnitude of carrier densities.
We identify the dominant FCA processes at different wavelengths, and the various investigated processes explain the fine structures in experimental measurements accurately.
Our results highlight the importance of considering both single-particle excitations including phonon-assisted and direct processes, as well as the semi-classical resistive contribution to obtain accurate FCA spectra.
Our method provides a tool to accurately evaluate the optical properties of doped semiconductors in practical optoelectronic devices. 
Our method is general, and can be extended to other material systems to quantitatively study the impact of FCA on their infrared applications, thus enables new oppotunities of the rational design of infrared optoelectronics.

\acknowledgments
The work is supported as part of the Computational Materials Sciences Program funded by the U.S. Department of Energy, Office of Science, Basic Energy Sciences, under Award No. DE-SC0020129.
Computational resources were provided by the National Energy Research Scientific Computing Center, which is supported by the Office of Science of the U.S. Department of Energy under Contract No. DE-AC02-05CH11231.

\section{Appendix}

\subsection{\label{sec:app_a}Detailed computational parameters}

In this section we provide detailed computational parameter for the calculations of Si optical properties in the main text. 
In our simulation, the wavefunctions are expanded into plane waves up to a cutoff energy of 60 Ryd.
To calculate quasiparticle energies in BerkeleyGW, the static dielectric function was calculated with a 30 Ry plane-wave cutoff and extended to finite frequency using the generalized plasmon-pole model of Hybertsen and Louie\cite{hybertsen1986}. 
The quasiparticle energies were determined on an 6$\times$6$\times$6 sampling grid of the first Brillouin zone(BZ).
The quasiparticle energies are subsequently interpolated using maximally localized Wannier function technique to a fine sampling grid of $48\times48\times48$ for free electrons and $80\times80\times80$ for free holes to determine the direct absorption with Fermi's golden rule. 
DFPT calculations are performed on a $6\times6\times6$ BZ sampling grid of phonon $\mathbf{q}$ vectors in the first BZ.
To calculate phonon-assisted,  charged-impurity-assisted absorption, and resistive contribution, the quasiparticle energies, the electron-phonon matrix elements and the phonon frequencies are interpolated to a fine sampling grid of $48\times48\times48$. 
For the phonon-assisted contribution, the second-order perturbation theory fails at ranges where first-order transitions are possible, and nonphysical divergence due to zeros in the denominators would be observed. 
Here we follow the approach of Brown et al.,\cite{doi:10.1021/acsnano.5b06199} to eliminate the divergence with a correction of $\varepsilon_2(\omega)=2\times\varepsilon_2(\omega)|_{2\eta}-\varepsilon_2(\omega)|_{\eta}$, using $\eta=0.01$ eV.
In order to calculate the absorption coefficient $\alpha(\omega)$ using the complex dielectric function, a constant value of the refractive index is assumed ($n_r=3.4$) thus $\alpha=\omega/c n_r \varepsilon(\omega)$, where $c$ is the speed of light. In the region below or close to the indirect band gap, this is a reasonable assumption due to the relatively negligible Im $\varepsilon(\omega)$ and the negligible effects of free carriers on Re $\varepsilon(\omega)$ and the refractive index $n_r$. 
In order to calculate the reference carrier mobilities for the resistive contribution, charged-impurity scatterings are taken into account using the point-charge model described in the main text.
The fine sampling of the electronic $\mathbf{k}$-grids and the phonon $\mathbf{q}$-grids are $100\times100\times100$ for both free electrons and free holes. 
Spin-orbit coupling is considered to calculate hole mobilities to account for the splitting of the valence band maximum. 

\subsection{\label{sec:app_b}Reference mobility}

\begin{figure}[!ht]
    \centering
    \includegraphics[width=0.5\textwidth]{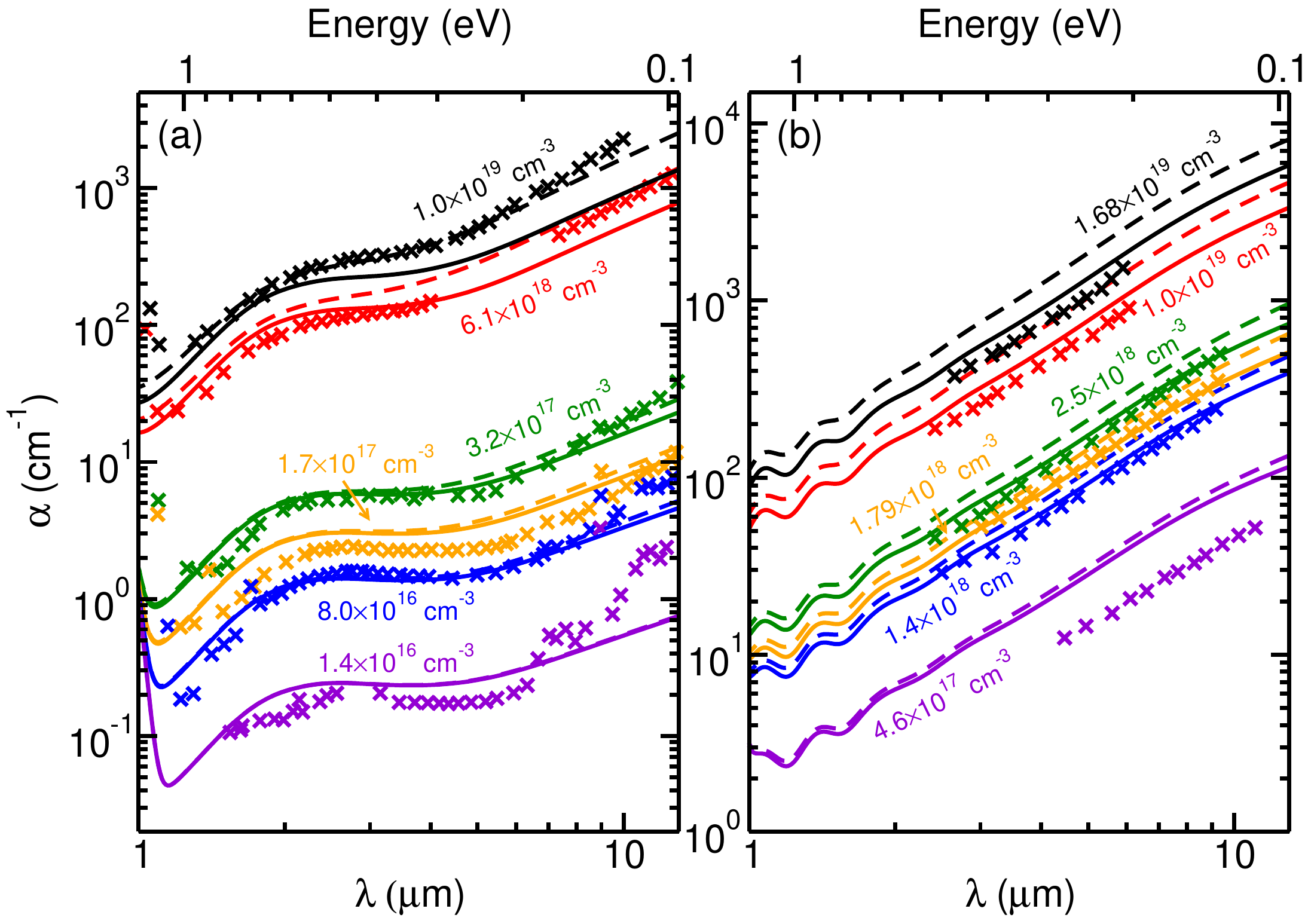}
    \caption{Calculated total absorption coefficient for (a) free electrons and (b) free holes with the resistive contribution evaluated from calculated mobility data (solid) or from experimental mobility data from Ref.\onlinecite{van2004principles} (dashed). 
    The reference mobility values affect calculated absorption coefficient quantitatively within a factor of two, but does not affect the qualitative conclusion. }
    \label{fig:ref_mob}
\end{figure}

In order to calculate the resistive contribution, a reference DC-conductivity is needed to obtain the relaxation time. 
In this work, the electron and hole mobilities of silicon were calculated at room temperature as a function of carrier concentration by iteratively solving the Boltzmann transport equation\cite{Ponce_2020}.  
The calculated electron mobilities agree well with experimental measurements from Ref.\cite{Jacobini:1977} 
at low carrier concentrations.
However, at the two highest carrier concentrations, theoretical calculations overestimate the electron mobilities compared to experimental measurements by a factor of 2.5.
This overestimation is mainly due to the lack of considering electron-electron interactions\cite{PhysRevB.94.115208}. 
The hole mobilities are overall overestimated by a factor of 2 $\sim$ 2.4 due to the effective masses of the heavy holes, which has been reported in previous studies\cite{PhysRevB.97.121201}. 
This overestimation of effective mass is consistent in calculating the mobility and the resistive contribution for the free-carrier absorption. 
We compared the absorption calculated from theoretical mobility values and from experimental mobility values from Ref.\onlinecite{van2004principles} and we show the comparison in Fig.\ref{fig:ref_mob}. We note that the choice of the mobility data does not affect the qualitative comparison and explanation of the shape of the absorption curve. 
Quantitatively, the absorption coefficient of free-electron absorption at high carrier concentrations, e.g., 10$^{19}$ cm$^{-3}$ is underestimated by a factor of 2.5 at wavelengths longer than 10 $\mu$m using theoretical data due to the overestimation of the mobility. 
The absorption coefficient for free-hole absorption is in better agreement consistently (with a maximum difference of 25\%) due to consistent effective masses when using theoretical data for the mobility. 

\subsection{\label{sec:app_c}Short-circuit current density}

\begin{figure*}[!ht]
    \hspace{-0.3in}
    \includegraphics[width=0.7\textwidth]{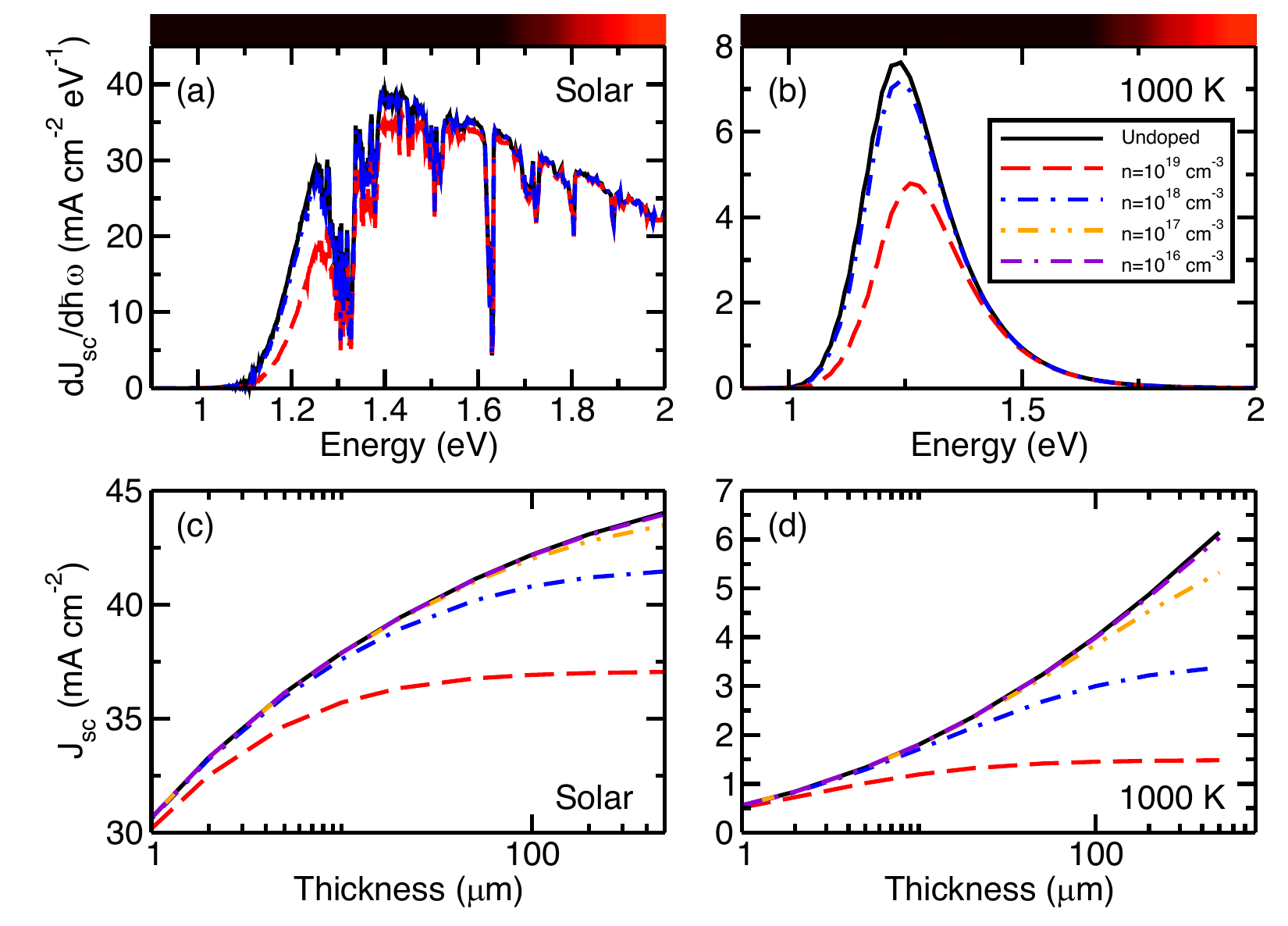}
    \vspace{-0.15in}
    \caption{(Color Online.) Effects of FCA on the performances of Si in optoelectronic devices. Calculated short-circuit current density per photon energy ($dJ_{sc}/d\hbar\omega$) from photon-induced electron-hole pairs as a function of photon energy for a Si thin film of 10 $\mu$m under (a) the AM1.5 solar spectrum\cite{solar} and (b) the ideal black-body spectrum at 1000 K. Integrated short-circuit current density ($J_{sc}$) as a function of film thickness for (c) the AM1.5 solar spectrum and (d) the ideal black-body radiation at 1000 K.
    The effects of FCA for electron densities of 10$^{17}$ cm$^{-3}$ and 10$^{16}$ cm$^{-3}$  are negligible in (a) and (b) thus not shown. 
    FCA does not have a strong detrimental effect on the performance of solar cells for typical cell thicknesses and carrier concentrations, but can affect thermal applications significantly. 
    }
    \label{fig:isc}
    \vspace{-0.1in}
\end{figure*}

In this section, we show more detailed analysis of the calculated current density for both the AM1.5 solar spectrum and the theoretical black-body spectrum of 1000 K. 
In Fig.\ref{fig:isc}(a) and (b), we show the calculated current density as a function of photon energy ($dJ_{sc}/d\hbar\omega$) for a 10 $\mu m$ thick film with various electron densities for both solar spectrum and the spectrum of ideal black body radiation at 1000 K, respectively. 
(c) and (d) show the corresponding integrated value ($J_{sc}$) for various film thicknesses. 
For solar cell applications, Fig.\ref{fig:isc}(a) shows that FCA only induces a very slight decrease of the short-circuit current close to the onset for a 10 $\mu$m Si thin-film with electron density of $10^{18}$ cm$^{-3}$.
For the integrated current density at $n=10^{18}\text{ cm}^{-3}$, the drop in $J_{sc}$ due to free electrons is above 1\% only for films thicker than 10 $\mu m$.
For carrier densities below $10^{17}\text{ cm}^{-3}$, the effects are negligible for films as thick as 100 $\mu m$. A typical solar cell based on Si has a very thin $n$-type emitter ($L\sim$1 $\mu m$)\cite{uprety2019photogenerated} and very low carrier concentration in the $p$-type ($n<$10$^{16}$ cm$^{-3}$) base. 
As illustrated in the figure, in these regions free-carrier absorption are negligible source of loss. 
Compared to the solar spectrum, the theoretical spectrum at 1000 K is significantly shifted towards the IR region. 
As a result, the strong effects of free carriers from 1 to 1.4 eV become more pronounced to the integrated current density.

\bibliography{ref}



\clearpage

\end{document}